\documentclass[11pt,a4]{article}
\usepackage{epsfig,graphics}
\usepackage{caption}
\begin{document}
\title{Correlations in nuclear energy recurrence relations}
\author{B. Buck, A.C. Merchant \\ 
Department of Physics, University of Oxford, Theoretical Physics, \\
1 Keble Road, Oxford OX1 3NP, UK. \\ and \\
S.M. Perez \\
Department of Physics, University of Cape Town, Private Bag, \\
Rondebosch 7700, South Africa. \\
and iThemba LABS, P.O.Box 722, Somerset West 7129, South Africa.
}%
\bigskip
\baselineskip 25.5pt
\maketitle
\bigskip
\begin{abstract}
\large
The excitation energies of states belonging to the ground state bands
of heavy even-even nuclei are analysed using recurrence relations.
Excellent agreement wih experimental data at the 10 keV level is
obtained by taking into account strong correlations which emerge in
the analysis. This implies that the excitation energies can be written
as a polynomial of maximum degree four in the angular momentum.
\end{abstract}
\centerline{21.10.-k, 21.10.Hw, 23.20.Lv and 27.20.+n} 
\section{Introduction}
\noindent
Despite, or perhaps because of the uncertainties in representing
the strong interaction by a potential, it has been a 
longstanding quest in nuclear physics to find simple
algebraic relationships between the excitation 
energies of a nucleus wherever possible. Attempts to achieve
this have often been inspired by approximate symmetries such
as Wigner's supermultiplets \cite{[Wigner]} or Elliot's SU(3)
scheme \cite{[Elliot]}. A textbook example leading to a simple
energy formula is provided by the rotational-vibrational 
model \cite{[BM],[Eichler]}. Later, the Interacting Boson Model
\cite{[AI]} used linear combinations of Casimir operators of
various chains of subgroups of U(6) to generate simple expressions 
for nuclear spectra. Indeed, investigations into algebraic cluster
models of energy levels continue to provide fruitful fields of 
research to the present day \cite{[Cseh]}.

However, the symmetries invoked by these models are never perfectly
realised in real nuclei, and although suggestively close agreement
can often be obtained in a number of favourable cases, they
generally struggle to achieve precision at the level of a
few 10's of keV. We suggest that it might be profitable to 
analyse the experimental data more closely without any
preconceived ideas about underlying models, to see if this alone
suggests any simple relations between excitation energies that 
could be used for predictive purposes. There have already been such
suggestions from our earlier work \cite{[BMP_2010]} and 
encouraged by these findings we examine here the deviations 
of our previous expressions from measured values with a view 
to improving their precision without sacrificing their simplicity.

It will turn out that the experimental excitation energies of five
consecutive states of the ground state band must be known for
our recurrence relations to get started. Whenever this applies,
they will provide a useful tool for predicting and/or confirming
the energies of the succeeding states of that band, and thus will
facilitate the task of nuclear data evaluation. We believe they should
be superior to the common practice of seeking to fit the energies with
a rotationally inspired formula such as 
$E = E_0 + aJ(J+1) +b[J(J+1)]^2$,  since they are based purely on
empirical values and are completely independent of any
underlying model. In particular, they will be helpful in predicting
the energy of the next state in a band beyond the highest identified 
to date.

\section{Recurrence relations}
We have previously analysed recursion relations involving
the angular momenta $J=0, 2, 4, \ldots$ and corresponding 
excitation energies $E(J)$ of states belonging to the ground 
state bands of even-even nuclei \cite{[BMP_2010]}. Given a
pair of angular momenta $(J,K)$ with $J>K$ we found empirically that
\begin{equation}
{E(J-L) - E(K+L) \over E(J)-E(K)} - {(J-L)-(K+L) \over (J-K)} = 0
\end{equation}
where $L$ increases in steps of two from an initial value $L=0$.
We note that as the spin values $(J-L)$ and $(K+L)$ change with 
$L$ their sum remains equal to $(J+K)$.

In the earlier analysis \cite{[BMP_2010]} we found that Eq.(1) was
satisfied to within a few percent with the deviations from that 
equation strongly correlated. These deviations are given by
\begin{equation}
d^{J,K}_L = {E(J-L) - E(K+L) \over E(J)-E(K)} - {(J-L)-(K+L) \over (J-K)}
\end{equation}
and here we investigate the nature and extent of the correlations 
between the various $d^{J,K}_L$, and how they can be used to improve
the agreement between theory and data. 

\section{Application to 87 heavy nuclei}
In the database shown in Table 1 \cite{[BNL_Data]} we have included all 87 
even-even nuclei between Ba and Cm which have more than four
protons (or proton-holes) and four neutrons (or neutron-holes)
outside closed shells, and for which the ground state bands are
accurately known (i.e. with firm $J^{\pi}$ assignments) up to at 
least $J^{\pi}=12^+$. 

We concentrate on the typical case defined by $J=10, K=0$ and 
$L_1=2, L_2=4$ leaving a more complete analysis to be presented 
elsewhere \cite{[BMP_Coming]}. Figure 1 shows that the 
deviations are surprisingly well described by the simple relation
\begin{equation}
d^{10,0}_2 = 2 \ d^{10,0}_4.
\end{equation}
Indeed, a least squares straight line fit to all 87 data points yields
a gradient of $2.007 \pm 0.028$ and an intercept of 
$0.00006 \pm 0.00032$. The Figure also indicates that the
six points with the highest values of $d_2^{10,0}$ deviate from the 
straight line in a statistically significant manner. These points
correspond to the nuclei $^{130}$Ce, $^{134}$Ce, $^{134}$Nd,
$^{136}$Sm, $^{140}$Gd and $^{172}$Os. The results for $^{134}$Ce
are probably problematic due to backbending. There are two $8^+$ 
states in close proximity at 2.81111 and 3.0176 MeV (we take the first
of them as the band member) and similarly three $10^+$ states at
3.2086, 3.7193 and 3.81765 MeV (we take the second of these as
the band member). 

This band crossing feature is generally expected for nuclei with
low deformation at the beginning and end of the rare-Earth region
\cite{[HeraSum]}. It therefore comes as no surprise that similar
deviations, albeit to a lesser extent,  are observed for four 
other nuclei in the neighbourhood of $^{134}$Ce. Table 2 lists the
states of the ground state band, and an excited band which can
be expected to mix with it, for all six of these outlying nuclei. We regard these 
few deviations in a positive light as pinpointing unusual nuclear 
spectra that are worthy of more detailed scrutiny, and here we 
would include also $^{150}$Sm, $^{174}$Os, $^{222}$Th and in
particular $^{244}$Pu.

The framework introduced above can be used to locate missing members 
of a band, or to check experimental assignments to the band. We
illustrate the procedure for the case where all the band members up
to $J=8$ are known and an estimate of the excitation energy of the
$J=10$ state is required. By Eq.(2)
\begin{equation}
d^{10,0}_2 = { E(8) - E(2) \over E(10) - E(0) } - 0.6 \ {\rm and}  \ 
 \ d^{10,0}_4 = {E(6) - E(4) \over E(10) - E(0) } - 0.2,
\end{equation}
which together with Eq.(3) result in 
\begin{equation}
\{ E(10) - E(0) \} = 5 \{ E(8) - 2E(6) + 2E(4) - E(2) \}
\end{equation}

Table 1 lists the individual ratios $r = d_2^{10,0}/d_4^{10,0}$ as well
as the differences $\Delta E$ between the theoretical
values of $\{ E(10) - E(0) \}$ given by Eq.(5) and their experimental
counterparts. Omitting the clearly anomalous case of $^{134}$Ce
we find an r.m.s. deviation of $\Delta E = 10.4$ keV, an
order of magnitude better than the results obtained when setting
$d_L^{J,K} = 0$ in either of Eqs.(4). Of interest also is that the further analysis undertaken below 
suggests that Eq.(5) is a particular case of a more general recurrence 
relation given by
\begin{equation}
\{ E(n) - E(n-10) \} = 5\{E(n-2) - 2E(n-4) + 2E(n-6) - E(n-8)\}
\end{equation}
with $n \geq  10$.

\section{Derivation from Taylor series}
This energy recurrence relation can also be obtained by assuming that the
excitation energy $E(n)$ may be expanded as a Taylor series in angular
momentum $n$ which converges sufficiently rapidly that fifth order and
higher terms may be neglected. Expanding $E(n_0 \pm 1)$, $E(n_0 \pm 3)$ and
$E(n_0 \pm 5)$ about $n_0$ we obtain
\begin{equation}
E(n_0 \pm 1) = E(n_0) \pm E^{\prime}(n_0) + {E^{\prime\prime}(n_0)
  \over 2!} \pm {E^{\prime\prime\prime}(n_0) \over 3!} + \ldots
\end{equation}
\begin{equation} E(n_0 \pm 3) = E(n_0) \pm 3E^{\prime}(n_0) + {9E^{\prime\prime}(n_0)
  \over 2!} \pm {27E^{\prime\prime\prime}(n_0) \over 3!} + \ldots
\end{equation}
\begin{equation}
E(n_0 \pm 5) = E(n_0) \pm 5E^{\prime}(n_0) + {25E^{\prime\prime}(n_0)
  \over 2!} \pm {125E^{\prime\prime\prime}(n_0) \over 3!} + \ldots
\end{equation}
By successively subtracting the $(n_0 - 1),\ (n_0 - 3)$ and $(n_0 - 5)$
terms from their $(n_0 + 1),\ (n_0 + 3)$ and $(n_0 + 5)$
counterparts in the three equations above we
completely eliminate all even powers from the Taylor series expansions
obtaining respectively,
\begin{equation}
[E(n_0+1)-E(n_0-1)] = 2E^{\prime}(n_0) + {E^{\prime\prime\prime}(n_0)
  \over 3} + \ldots
\end{equation}
\begin{equation}
[E(n_0+3)-E(n_0-3)] = 6E^{\prime}(n_0) + 9E^{\prime\prime\prime}(n_0)
 + \ldots
\end{equation}
\begin{equation}
[E(n_0+5)-E(n_0-5)] = 10E^{\prime}(n_0) + {250E^{\prime\prime\prime}(n_0)
  \over 6} + \ldots
\end{equation}
Subtracting twice Eq.(10) from Eq.(11) yields
\begin{equation}
E(n_0+3) - E(n_0-3) - 2E(n_0+1) + 2E(n_0-1) = 2E^{\prime}(n_0) + {50 \over 6}E^{\prime\prime\prime}(n_0) + \dots
\end{equation}
and the right hand side of Eq.(12) is exactly
five times that of Eq.(13). This allows us to write
\begin{equation}
\{ E(n_0+5) - E(n_0-5) \} = 5\{E(n_0+3) - 2E(n_0+1) + 2E(n_0-1) - E(n_0-3)\}
\end{equation}
so that taking $n_0=5$ we obtain Eq.(5) which, taken
together with Eq.(4), yields Eq.(3). Also by putting  
$n_0=n-5$ in Eq.(14) we recover Eq.(6).  We thus find that Eq.(6),
which with $n=10$ generates the excellent agreement
with the data shown in Table 1, can be derived from a
Taylor series expansion if terms of order five and higher
are ignored. 

Of course, we have not rigorously proved convergence of the Taylor
series, which depends on the (unknown) behaviour of the higher derivates
of $E$ as a function of $J$. What we have done, starting from a Taylor
series, is to completely eliminate all even powers of $J$, and show that 
Eqs.(3), (5) and (6) follow if the remaining odd terms of order 
5 and higher are ignored. This suggests that the band excitation energies $E(J)$
considered here take the form of a polynomial of maximum degree four in the 
angular momentum $J$, and we note that only cubic forms
were considered in earlier work \cite{[BMP_2010],[ZamCas]}.

Since Eq.(6) is satisfied by any E-dependence on $J$ that is of
quartic or lower degree, it encompasses a number of models.
It is satisfied, for example, by the idealised quadratic form 
$E_0 + \alpha J(J+1)$ of a perfect rotor with a constant moment of 
inertia, or by the anharmonic vibrator (AHV) model for which
$E(J) = \epsilon_2 n + \epsilon_4 n(n-1)/2 + \epsilon_6 n(n-1)(n-2)/6$,
where $n=J/2$ and $\epsilon_2, \epsilon_4$ and $\epsilon_6$ are
parameters fitted to each nucleus in turn \cite{[ZamCas]}.
We note, however, that although Eq.(6) is obeyed by these models
it is satisfied to the extent that inserting the model quantities on
the right hand side of Eq.(6) generates the model quantity on the
left hand side. As stated in our introductory paragraph, our 
emphasis is to extend previous work \cite{[BMP_2010]} by
finding more accurate {\it direct\/} relations between
{\it experimental\/} quantities. 

\section{Conclusion}
In conclusion we have found strong correlations between the deviations
$d^{J,K}_L$ from our original relations of Eq.(1) above. This has
enabled us to write down a highly successful recursion relation,
not involving any free parameters, for locating members of 
ground state bands from a knowledge of the excitation energies 
of the lower members of the band. For the specific case of $J=10, K=0$ 
and $L_1=2, L_2=4$ we have shown how a knowledge of the excitation
energies of the $J=2, 4, 6, 8$ band members allows us to predict
the excitation energy of the $J=10$ state with an r.m.s. error of
$\sim 10$ keV. We note that this is likely to be a particular case of a more
general relation. Although we have restricted attention here to values of $J$ no 
higher than 10, we have observed that Eq.(6) can be used 
to predict energies for states with much greater
values of $J$ (up to $J \sim 30$ in some cases) and we intend to
investigate this further elsewhere.

\vskip 1truecm
\noindent
{\bf Acknowledgements:} \ \
One of us (S.M.P.) acknowledges partial support for this work
from the National Research Foundation of South Africa under
grant UID 85382. The opinions, findings and conclusions 
expressed in any publication generated by NRF supported
research are those of the authors, and the NRF accepts no
liability whatsoever in this regard.

\begin{figure}
\vspace{0mm}
\hspace{0mm}
\includegraphics{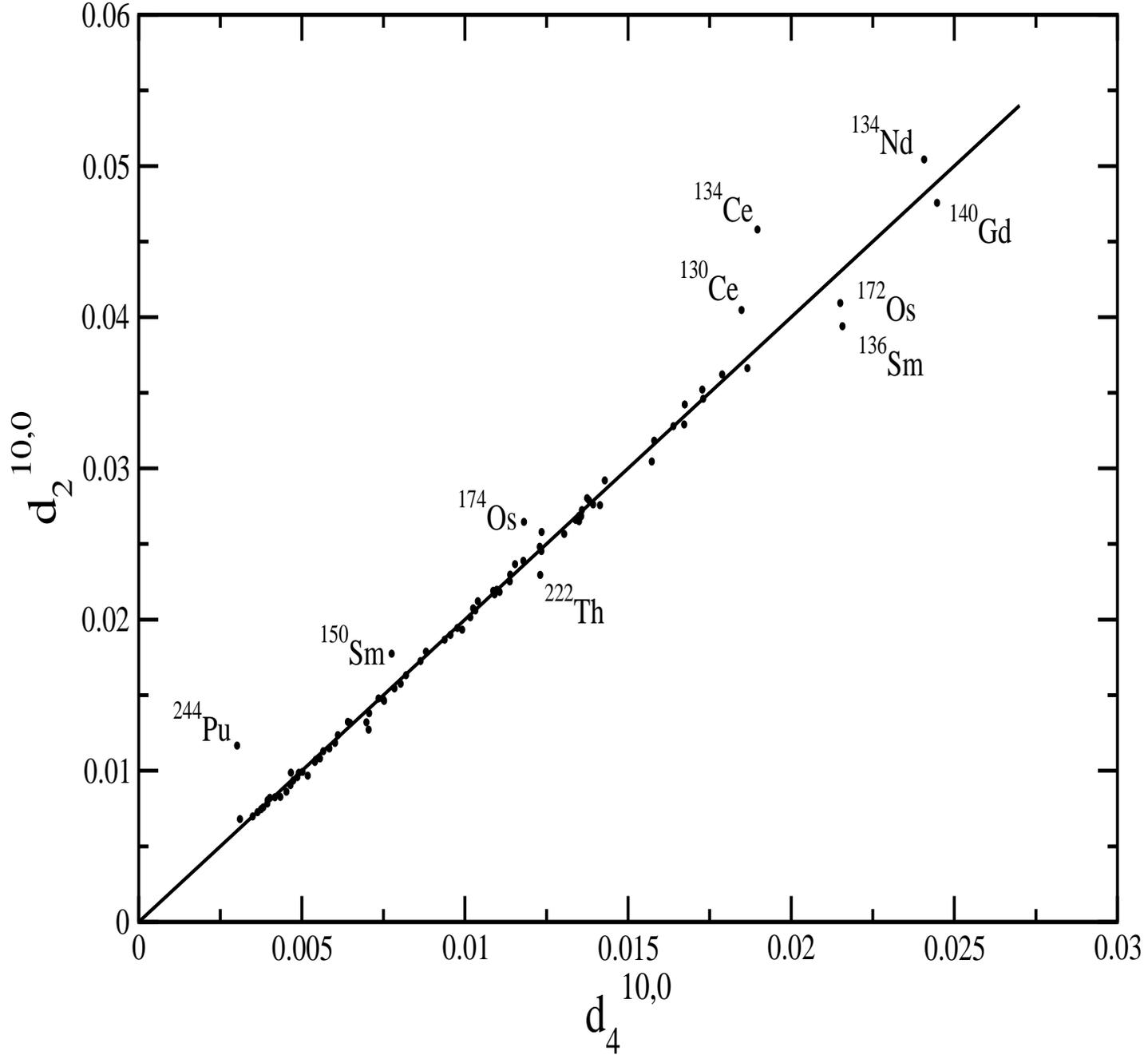}
\vspace{0mm}
\caption{Plot of the deviations $d_2^{10,0}$ against the deviations 
$d_4^{10,0}$ for the 87 nuclei listed in Table 1. A least squares 
straight line fit to the data indicates a slope of $2.007 \pm 0.028$
and an intercept of $0.00006 \pm 0.00032$.
The solid line corresponds to $d_2^{10,0} = 2d_4^{10,0}$.} 
\vspace{0mm}
\end{figure}

\begin{center}
\begin{table}[htb]
\caption{Database of 87 rare-Earth and Actinide nuclei.
We highlight the
ratios $r=d_2^{10,0}/d_4^{10,0}$ from Eq.(4), which are all close to
2.0, and $\Delta E$ the difference between the values of the 
$10^+$ excitation energy derived using Eq.(5) and that
obtained from direct measurement.}
\smallskip
\begin{tabular}{lccccrrrrr}
\hline
\smallskip
Table & \multicolumn{4}{c}{Nucleus} & $d_2^{10,0}$ & $d_4^{10,0}$ &
$r$ & $E(10^+)$ & $\Delta E$ \\
Position &  & $Z$ & $N$ & $A$ & & & & (keV) & (keV) \\
\hline
01&Ba&56&72&128&0.017882&0.008805&2.030&3082.3&4.199\\
02&Ba&56&88&144&0.021973&0.010976&2.001&2044.3&0.200\\
03&Ba&56&90&146&0.034223&0.016736&2.044&2051.8&7.699\\
04&Ce&58&66&124&0.022971&0.011385&2.017&2100.9&2.099\\
05&Ce&58&72&130&0.040477&0.018476&2.190&2809.0 &49.49\\
06&Ce&58&76&134&0.045793&0.018965&2.414&3719.3&146.2\\
07&Ce&58&92&150&0.022522&0.011373&1.980&1422.6&-1.60\\
08&Nd&60&74&134&0.050431&0.024076&2.094&2816.9&32.10\\
09&Nd&60&94&154&0.010588&0.005401&1.960&1210.8&-1.29\\
10&Nd&60&96&156&0.008212&0.004020&2.042&1169.0&1.000\\
11&Sm&62&72&134&0.027561&0.014139&1.949&1952.0&-7.00\\
12&Sm&62&74&136&0.039401&0.021568&1.826&2414.6&-45.1\\
13&Sm&62&88&150&0.017746&0.007751&2.289&2433.2&27.30\\
14&Sm&62&90&152&0.023663&0.011533&2.051&1609.2&4.799\\
15&Sm&62&92&154&0.015753&0.008027&1.962&1333.0&-2.00\\
16&Sm&62&94&156&0.008612&0.004528&1.902&1307.4&-2.89\\
17&Sm&62&98&160&0.007835&0.003942&1.987&1227.8&-0.29\\
18&Gd&64&74&138&0.030454&0.015726&1.936&2266.3&-11.3\\
19&Gd&64&76&140&0.047561&0.024470&1.943&2796.8&-19.2\\
20&Gd&64&88&152&0.009676&0.005181&1.867&2300.4&-7.90\\
21&Gd&64&90&154&0.023885&0.011789&2.025&1637.0&2.500\\
22&Gd&64&92&156&0.018670&0.009377&1.990&1416.1&-0.59\\
23&Gd&64&94&158&0.010814&0.005555&1.946&1350.0&-2.00\\
24&Gd&64&96&160&0.009364&0.004735&1.977&1300.7&-0.70\\
25&Gd&64&98&162&0.009580&0.004863&1.970&1237.9&-0.89\\
26&Dy&66&88&154&0.013211&0.006978&1.893&2304.1&-8.59\\
27&Dy&66&90&156&0.024811&0.012289&2.018&1725.0&2.000\\
28&Dy&66&92&158&0.021669&0.010907&1.986&1520.1&-1.10\\
29&Dy&66&94&160&0.016316&0.008193&1.991&1428.0&-0.49\\
30&Dy&66&96&162&0.011300&0.005657&1.997&1375.1&-0.09\\
31&Dy&66&98&164&0.010719&0.005422&1.976&1261.3&-0.79\\
32&Dy&66&100&166&0.008053&0.003952&2.037&1341.0&0.999\\
33&Er&68&88&156&0.013231&0.006410&2.063&2633.1&5.400\\
34&Er&68&90&158&0.027889&0.013799&2.020&2072.5&2.999\\
35&Er&68&92&160&0.026597&0.013389&1.986&1761.1&-1.60\\
36&Er&68&94&162&0.020601&0.010319&1.996&1602.8&-0.29\\
37&Er&68&96&164&0.014715&0.007496&1.963&1518.1&-2.09\\
38&Er&68&98&166&0.015441&0.007839&1.969&1349.6&-1.59\\
39&Er&68&100&168&0.007459&0.003751&1.988&1396.8&-0.29\\
40&Yb&70&90&160&0.029203&0.014285&2.044&2373.0&7.499\\
41&Yb&70&92&162&0.031836&0.015799&2.015&2024.1&2.400\\
42&Yb&70&94&164&0.027238&0.013585&2.005&1753.4&0.600\\
43&Yb&70&96&166&0.020150&0.010162&1.982&1605.9&-1.40\\
44&Yb&70&98&168&0.018984&0.009555&1.986&1425.4&-0.90\\
45&Yb&70&100&170&0.011478&0.005843&1.964&1437.5&-1.50\\
\hline
\end{tabular}
\end{table}
\end{center}
\begin{center}
\begin{table}[htb]
\caption*{Database of 87 rare-Earth and Actinide nuclei continued.}
\smallskip
\begin{tabular}{lccccrrrrr}
\hline
\smallskip
Table & \multicolumn{4}{c}{Nucleus} & $d_2^{10,0}$ & $d_4^{10,0}$ &
$r$ & $E(10^+)$ & $\Delta E$ \\
Position &  & $Z$ & $N$ & $A$ & & & & (keV) & (keV) \\
\hline
46&Hf&72&90&162&0.028026&0.013743&2.039&2635.4&7.100\\
47&Hf&72&94&166&0.032790&0.016390&2.000&1971.9&0.100\\
48&Hf&72&96&168&0.027613&0.013927&1.982&1736.1&-2.09\\
49&Hf&72&100&172&0.019445&0.009768&1.990&1521.2&-0.69\\
50&Hf&72&106&178&0.014786&0.007348&2.012&1570.3&0.699\\
51&W&74&106&180&0.007580&0.003815&1.987&1630.4&-0.39\\
52&W&74&94&168&0.036210&0.017882&2.024&2202.1&4.899\\
53&W&74&96&170&0.034604&0.017302&2.000&1901.5&2.220\\
54&W&74&98&172&0.032906&0.016719&1.968&1617.3&-4.29\\
55&W&74&100&174&0.026503&0.013496&1.963&1637.5&-3.99\\
56&W&74&102&176&0.025659&0.013042&1.967&1648.5&-3.50\\
57&W&74&104&178&0.021832&0.011060&1.973&1665.4&-2.40\\
58&W&74&106&180&0.021897&0.010864&2.015&1664.1&1.400\\
59&W&74&108&182&0.009929&0.005023&1.976&1712.0&-0.99\\
60&W&74&110&184&0.013177&0.006470&2.036&1860.8&2.200\\
61&Os&76&96&172&0.040940&0.021503&1.903&2023.9&-20.9\\
62&Os&76&98&174&0.026460&0.011808&2.240&1617.5&23.00\\
63&Os&76&100&176&0.025787&0.012349&2.088&1634.1&8.899\\
64&Os&76&104&180&0.036625&0.018658&1.963&1767.6&-6.09\\
65&Os&76&106&182&0.035209&0.017273&2.038&1812.0&5.999\\
66&Os&76&108&184&0.017250&0.008636&1.997&1871.2&-0.20\\
67&Os&76&110&186&0.020744&0.010251&2.023&2068.0&2.499\\
68&Os&76&116&192&0.021217&0.010393&2.041&2418.8&5.200\\
69&Ra&88&132&220&0.012720&0.007045&1.805&1342.7&-9.19\\
70&Ra&88&138&226&0.026836&0.013563&1.978&959.9&-1.39\\
71&Th&90&132&222&0.022955&0.012305&1.865&1461.1&-12.1\\
72&Th&90&134&224&0.026852&0.013494&1.989&1173.8&-0.79\\
73&Th&90&136&226&0.024531&0.012342&1.987&1040.3&-0.79\\
74&Th&90&138&228&0.019324&0.009914&1.949&911.8&-2.29\\
75&Th&90&142&232&0.013812&0.007063&1.955&826.8&-1.30\\
76&U&92&138&230&0.014621&0.007520&1.944&856.3&-1.79\\
77&U&92&140&232&0.012358&0.006104&2.024&805.9&0.600\\
78&U&92&142&234&0.011845&0.006017&1.968&741.2&-0.69\\
79&U&92&144&236&0.009868&0.004908&2.010&782.3&0.199\\
80&U&92&146&238&0.009872&0.004665&2.116&775.9&2.100\\
81&Pu&94&142&236&0.009049&0.004654&1.944&773.5&-1.0\\
82&Pu&94&144&238&0.006981&0.003490&1.999&773.5&-3.33\\
83&Pu&94&146&240&0.008241&0.004174&1.974&747.4&-0.40\\
84&Pu&94&148&242&0.008271&0.004341&1.905&778.6&-1.60\\
85&Pu&94&150&244&0.011665&0.003015&3.867&802.4&22.6\\
86&Pu&94&152&246&0.007260&0.003642&1.993&818.1&-0.10\\
87&Cm&96&152&248&0.006809&0.003102&2.194&760.7&2.300\\
\hline
\end{tabular}
\end{table}
\end{center}

\begin{center}
\begin{table}[htb]
\caption{Excitation energies (keV) of ground state and nearby excited
state band members in the six outlying nuclei $^{130}$Ce, 
$^{134}$Ce, $^{134}$Nd, $^{136}$Sm, $^{140}$Gd and $^{172}$Os. 
The superscripts are the band labels used in the Evaluated
Nuclear Structure Data File \cite{[BNL_Data]}.}
\smallskip
\begin{tabular}{lrrrrrr}
\hline
\smallskip
$J^{\pi}$ & $^{130}$Ce & $^{134}$Ce & $^{134}$Nd & $^{136}$Sm &
$^{140}$Gd & $^{172}$Os \\
\hline
$2^+$ & 253.85$^e$ & 409.20$^\&$ & 294.17$^\&$ & 254.92$^\&$ &  
328.6$^@$ & 227.77$^a$ \\
$4^+$ & 710.37$^e$ & 1048.68$^\&$ & 788.92$^\&$ & 686.36$^\&$ &  
836.2$^@$ & 606.17$^a$ \\
$6^+$ & 1324.1$^e$ & 1863.1$^\&$ & 1420.06$^\&$ & 1221.4$^\&$ &  
1464.0$^@$ & 1054.47$^a$ \\
$8^+$ & 2053.1$^e$ & 2811.1$^\&$ & 2126.4$^\&$ & 1798.8$^\&$ &  
2139.7$^@$ & 1524.95$^a$ \\
$10^+$ & 2809.0$^e$ & 3719.3$^\&$ & 2816.9$^\&$ & 2414.6$^\&$ &  
2796.8$^@$ & 2023.87$^a$ \\
$12^+$ & 3311.9$^e$ & 4183.6$^\&$ & 3482.9$^\&$ & 3091.8$^\&$ &  
3267.5$^@$ & 2564.5$^a$ \\
& & & & & &  \\
$2^+$ &  834.55$^f$ & 965.66$^c$ & 793.86$^a$ & 712.88$^c$ &  
713.3$^\&$ & 810.01$^c$ \\
$4^+$ & 1322.83$^f$ & 1643.47$^c$ &1313.03$^a$ & 1170.98$^c$ &  
1281.4$^\&$ & 1137.88$^c$ \\
$6^+$ & 1897.6$^f$ & 2303.8$^c$ & 1910.6$^a$ & 1640.96$^c$ &  
1881.4$^\&$ & 1551.25$^c$ \\
$8^+$& 2560.6$^f$ & 3017.6$^c$ & 2467.2$^a$ & 2250.2$^c$ &  
2632$^\&$ & 2093.63$^c$ \\
$10^+$ & 3296.5$^f$ & 3208.6$^d$ & 3051.9$^a$ & 2953.7$^c$ &  
2926.8$^b$ & \\
$12^+$ & 3985.3$^f$ & 4006.8$^d$ & 3436.5$^a$ & 3682.6$^c$ &  
3617$^b$ & \\
\hline
\end{tabular}
\end{table}
\end{center}

\end{document}